\documentclass[11pt,a4paper]{article}
\usepackage[left=1in, right=1in, top=1in, bottom=1in]{geometry}
\usepackage{pgfplots}
\pgfplotsset{width=7cm,compat=1.15}
\usepackage{graphicx}
\usepackage{bm}
\newcommand*{\eqref}[1]{(\ref{#1})}
\usepackage[skip=8pt,font=scriptsize]{caption}
\begin{document}
\begin{center}
{\large{The electron-proton bound state in the continuum with the positive binding energy of 1.531 of the electron mass.}}\\

\vspace{.5cm}

A.I. Agafonov

\vspace{.5cm}

National Research Centre "Kurchatov Institute", Moscow 123182, Russia\\

Moscow Aviation Institute (National Research University), Moscow, 125993, Russia\\

\vspace{.5cm}

Agafonov\_AIV@nrcki.ru, aiagafonov7@gmail.com

\vspace{.5cm}

\end{center}
\abstract{
In the bound states in the continuum (BIC) the binding energy is positive, and the mass of a composite particle is greater than the total mass of its constituents. In this work the BIC
state is studied for the electron-proton system with using the ladder Bethe-Salpeter equation. We demonstrate that there are two momentum space regions in which the electromagnetic
interaction between the particles is strongly enhanced, and the effective coupling constant is equal to $\alpha \sqrt{m_{p}/m_{e}}=0.313$, where $\alpha$  is the fine structure constant, 
$m_{p}$ and $m_{e}$ are the proton and the electron masses. This interaction resonance causes the confinement of the pair in the BIC state with the positive binding energy of 1.531 of the electron mass. The integral equation for the bispinor wave function is derived. This normalized wave function which must be complex, was found numerically in the momentum and coordinate spaces. It turned out that in the BIC state, the average radius for the electron is equal to 48Fm, and the average radius for the proton is equal to 1.1Fm. This composite particle can exist
exclusively in the free state, in which its properties, such as its form-factors, should only be studied. In bound states with other particles, the composite loses its individuality.}

\vspace{.5cm}                             



\vspace{1cm}

\section{Introduction}

\par Unlike the conventional bound states, there are the bound states in the continuum (BIC) in which the binding energy is positive and the composite particle mass is greater than the total
mass of the constituents. The BIC states have been discovered by von Neumann and Wigner in 1929 \cite{bib1} (see also \cite{bib2} with some extension and correction of this work). 
It so happened that the BIC phenomenon was forgotten for about 40 years \cite{bib3}. Thereafter, the BIC states have been found experimentally in condensed matter physics and optics 
(see \cite{bib4,bib5,bib6,bib7,bib8} and references therein). These states are stable due to the confinement mechanisms that are individual for each case.

\par In the non-relativistic mechanics, the BIC states have been investigated by using the Schrödinger Hamiltonian. As a rule, the spectral analysis of the real equation $H\psi=E\psi$
with $E>0$ was carried out \cite{bib4,bib5,bib9}. However, the eigenvalues of the BIC states are in the continuous spectrum. Therefore, the study of the BIC states must be carried out with
using the Lippmann-Schwinger integral equation, in which an infinitesimal $\delta$ is added to the energy $E$. Then, the kernel of the integral equation and, respectively, the BIC wave functions become complex. As a result, the spectral analysis should be provided for a system of two coupled integral equations. 

\par As far as we know in quantum electrodynamics the BIC states have not been supposed and studied previously. Furthermore, particle physics has been developed without the analysis
of these states. The our idea is to apply this BIC phenomenon to some elementary particles. In the BIC state of the electron-proton system, if the state exists, the composite
boson mass $m_{B}>m_{p}+m_{e}$. In fact, the BIC state would be of interest if the positive binding energy ${\cal E}=m_{B}-m_{p}-m_{e}\simeq 1.531m_{e}$. 

\par The  coupling constant of the electromagnetic interaction is the fine structure constant, $\alpha=\frac{1}{137.04}$. Because of the relatively small value of the coupling constant, the 
mass of the composite particle $E$ in the normal bound states is slightly less than the mass of the constituent particles $\sum_{i}m_{i}$. Then the binding energy defined as 
$E-\sum_{i}m_{i}$ is negative, and, as a rule, is proportional to $\alpha^2$. Typical examples are the hydrogen and the positronium.

\par In this regard, the following questions arise: 1) How can an electromagnetic interaction with the relatively small coupling constant leads to composite particle with large, positive
value of the binding energy? 2) The composite from the electron and the proton is the boson with the spin equal to 0 or 1. Can the behavior of this boson in inhomogeneous
magnetic fields represents fermion properties? The form-factors of the composite boson can be extracted from the calculated wave function in the BIC state. Can these form-factors be
compared with the known data on the neutron form-factors?

\par These issues are discussed in the present paper. The composite particle of the proton and the electron is studied. Using the ladder Bethe-Salpeter equation, an integral equation for
the BIC state of the two-particle system is derived. We demonstrate that the electromagnetic interaction between the particles is strongly enhanced when the momenta of the constituents
are in the two regions in the momentum space. These momentum-space regions can be called the resonant regions because the interaction between particles becomes formally unlimited.
Together with the correlations in particles motion, the resonance of the electromagnetic interaction leads to the confinement of these particles in the BIC state with positive binding
energy equal to $1.531m_{e}$. Numerical solution of this integral equation by means of the iteration method is found. Results obtained for the BIC wave function in the momentum
and coordinate spaces, are presented.

\par Natural units ($\hbar=c= 1$) will be used throughout.

\section{The equal-time Bethe-Salpeter equation}
The bound states are described by the homogeneous Bethe-Salpeter equation \cite{bib10}:
\begin{equation}\label{1}
\psi(1,2)=-i\int\int\int\int d\tau_{3}d\tau_{4}d\tau_{5}d\tau_{6} K_{e}(1,3)K_{p}(2,4)G(3,4;5,6)\psi(5,6).
\end{equation}
In Eq. \eqref{1} $d\tau_{i}=d{\bf r}_{i}dt_{i}$, $K_{e}$ and $K_{p}$ are the free propagators for the electron an the proton, and $G(3,4;5,6)$ is the interaction function. In the ladder approximation the function is given by:
\begin{equation}\label{2}   
G^{(1)}(3,4;5,6)=-\alpha (1-{\bm \alpha}_{e}{\bm \alpha}_{p})\delta^{(4)}(3,5)\delta^{(4)}(4,6)\delta_{+}(s_{56}^2),
\end{equation} 
where $\delta_{+}(s_{56}^2)$ is the propagation function of the virtual photon.

\par For the problems of bound states, the free particle  propagator was discussed in \cite{bib11}. The electron propagator takes the form:
\begin{equation}\label{3}   
K_{e}(1,3)=\sum_{\bf p}\frac{1}{2\varepsilon_{p}}\Bigl[ \Lambda_{e}^{+}e^{-i\varepsilon_{p}(t_{1}-t_{3})}+\Lambda_{e}^{-} e^{i\varepsilon_{p}(t_{1}-t_{3})}\Bigr]\theta(t_{1}-t_{3})
e^{i{\bf p}({\bf r}_{1}-{\bf r}_{3})},
\end{equation} 
where
\begin{equation}\label{4}     
\Lambda^{\pm}_{e}({\bf p})=\varepsilon_{{\bf p}}\pm {\bm \alpha}_{e}{\bf p}\pm \beta_{e}m_{e},
\end{equation}
$m_{e}$, ${\bf p}$ and $\varepsilon_{{\bf p}}=\sqrt{m_{e}^2+p^2}$ are the mass, the momentum and the energy of the electron, and ${\bm \alpha}_{e}$ and $\beta_{e}$ matrices are taken in the standard representation.
\par Respectively, the propagator of the particle $p$ is given by:
\begin{equation}\label{5}   
K_{p}(2,4)=\sum_{\bf q}\frac{1}{2\omega_{q}}	\Bigl[ \Lambda_{p}^{+}e^{-i\omega_{q}(t_{2}-t_{4})}+\Lambda_{p}^{-} e^{i\omega_{q}(t_{2}-t_{4})}\Bigr]\theta(t_{2}-t_{4})
e^{i{\bf q}({\bf r}_{2}-{\bf r}_{4})}
\end{equation} 
Here 
\begin{equation}\label{6}
\Lambda^{\pm}_{p}({\bf q})=\omega_{\bf q}\pm {\bm \alpha}_{p}{\bf q}\pm \beta_{p}m_{b},
\end{equation}
$m_{p}$, ${\bf q}$ and $\omega_{{\bf q}}=\sqrt{m_{p}^2+q^2}$ are the mass, the momentum and the energy of the proton, and ${\bm \alpha}_{p}$ and $\beta_{p}$ matrices are taken in the standard representation.

\par We search the BIC state with the binding energy proportional to the mass of the lighter particle from the pair, ${\cal E}\propto m_{e}$. The energy of the virtual photon, $\omega$, is about of the binding energy. We suppose that in the BIC state, characterized scale of the distance between the two particles, $r_{56}$, is: 
\begin{equation}\label{7}
r_{56}<< \frac{m_{e}}{{\cal E}}{\mathchar'26\mkern-10mu\lambda}_{e},
\end{equation}
where ${\mathchar'26\mkern-10mu\lambda}_{e}$ is the Compton wavelength of the electron. 

\par  Obviously, \eqref{7} implies a physically reasonable size of the composite particle. Inequality \eqref{7} means that we can neglect the retardation of the interaction between the particles. Then, Eq. \eqref{1} is reduced to the equal-time Bethe-Salpeter equation:
\begin{equation}\label{8}
\psi({\bf r}_{1},{\bf r}_{2};E)=-\alpha \int d{\bf r}_{1}^{\prime} \int d{\bf r}_{2}^{\prime} \sum_{\bf p}\sum_{\bf q} 
K_{ep}({\bf p},{\bf q};E)e^{i{\bf p}({\bf r}_{1}-{\bf r}_{1}^{\prime})+i{\bf q}({\bf r}_{2}-{\bf r}_{2}^{\prime})}
\frac{1-{\bm \alpha}_{e}{\bm \alpha}_{p}}{\vert {\bf r}_{1}^{\prime}-{\bf r}_{2}^{\prime} \vert}
\psi({\bf r}_{1}^{\prime},{\bf r}_{2}^{\prime};E).
\end{equation}
Here the two-particle propagator is:
\begin{equation}\label{9}
K_{ep}=
\frac{1}{4\omega_{q}\varepsilon_{p}}
\Bigl[\frac{\Lambda_{p}^{+}\Lambda_{e}^{+}}{E-\varepsilon_{p}-\omega_{q}+i\delta}+\frac{\Lambda_{p}^{+}\Lambda_{e}^{-}}{E+\varepsilon_{p}-\omega_{q}+i\delta}
\frac{\Lambda_{p}^{-}\Lambda_{e}^{+}}{E-\varepsilon_{p}+\omega_{q}+i\delta}+\frac{\Lambda_{p}^{-}\Lambda_{e}^{-}}{E+\varepsilon_{p}+\omega_{q}+i\delta}
\Bigr]
\end{equation}

In the momentum space, Eq. \eqref{8} with \eqref{9} is reduced to the form:
\begin{equation}\label{10}
\psi({\bf p},{\bf q};E)=-\frac{\alpha}{2\pi^2} K_{ep}({\bf p},{\bf q};E)
\int \frac{d{\bf k}}{k^2}(1-{\bm \alpha}_{e}{\bm \alpha}_{p})\psi({\bf p}+{\bf k};{\bf q}-{\bf k}).
\end{equation}

\section{Resonance of the electromagnetic interaction}
Consider the BIC state with the total energy $E=m_{p}+m_{e}+{\cal E}$ and the binding energy ${\cal E}=1.531 m_{e}$.
The imaginary parts of the last two terms in the square brackets in Eq. \eqref{9} vanish. These terms are not essential for the formation of the BIC state and can be omitted. The first two terms in Eq. \eqref{9} are fundamentally important. Using them, in Eq. \eqref{10} we can introduce a function describing the effective interaction between the particles:    
\begin{displaymath}
\alpha_{eff}=\frac{\alpha\Lambda_{e}^{+}\Lambda_{p}^{+}}{4\omega_{q}\varepsilon_{p}}
\Bigl[\frac{\cal P}{E-\varepsilon_{p}-\omega_{q}}-i\delta(E-\varepsilon_{p}-\omega_{q})\Bigr]+
\end{displaymath}
\begin{equation}\label{11} 
\frac{\alpha\Lambda_{e}^{-}\Lambda_{p}^{+}}{4\omega_{q}\varepsilon_{p}}
\Bigl[\frac{\cal P}{E+\varepsilon_{p}-\omega_{q}}-i\delta(E+\varepsilon_{p}-\omega_{q})\Bigr]
\end{equation}

\par For normal bound states the energy eigenvalue is negative, ${\cal E}<0$. Then the $\delta$-functions in the right-hand side of Eq. \eqref{11} vanish as well. Hence any enhancement of the interaction between particles does not occur in the convenient bound states with the negative binding energy. 

\par For the BIC state the energy eigenvalue is positive, ${\cal E}>0$. Hence $E-m_{p}\pm m_{e}>0$, and the term with the $\delta$-function is principally important. This function determines  
the two regions in the momentum space which are solutions of the equation:
\begin{equation}\label{12}  
(E-\omega_{\bf q})^2=\varepsilon_{\bf p}^2.
\end{equation}
Inside these regions the principal part 
\begin{equation}\label{13}  
\frac{\cal P}{(E-\omega_{q})^2-\varepsilon_{\bf p}^2}=0.  
\end{equation}

\begin{figure}[ht]
\centering
\includegraphics[width=16cm]{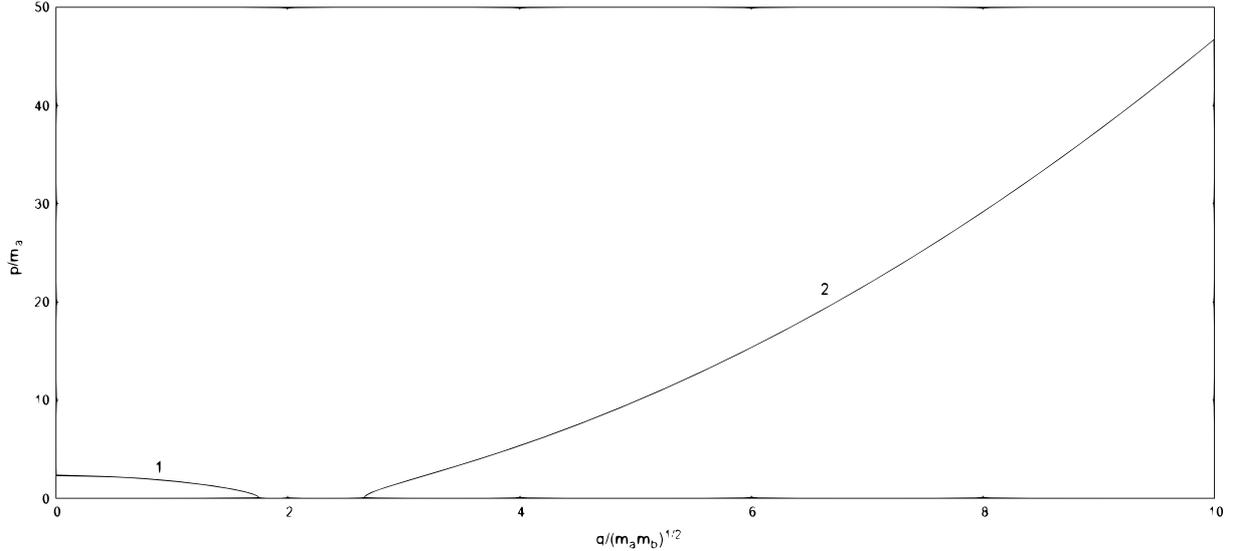}
\caption{The solution of Eq. \eqref{12} for the energy $E=m_{p}+m_{e}+{\mathcal E}$ with ${\mathcal E}= \gamma m_{e}$ and $\gamma=1.531$. 
\label{f1}}
\end{figure}
\par The two lines in Fig. 1 correspond  the two regions in the momenta space. These lines are given the relation between the modules of the vectors ${\bf p}$ and ${\bf q}$. Their directions, ${\bf p}/p$ and ${\bf q}/q$ are also interrelated, as will be shown below. In these regions, the motions of two particles are correlated with each other, and their interaction is sharply enhanced, becoming formally unlimited. Therefore, these regions can be called resonant ones.
 
\par The first resonant region is presented by curve 1 in Fig. 1. For the electron, this region is limited from above by the moment $p=2.324m_{e}$, and for the proton the similar restriction is $q\leq 1.749\sqrt{m_{e}m_{p}}$. Here the energies of the both particles are positive. This curve is given by the first term in square brackets on the right side of Eq. \eqref{9}. However, this region is not important for the formation of the BIC state that is of interest to us. This is because the electron momenta are relatively small. Respectively, it leads inevitably to large radius of the composite particle that is about ${\mathchar'26\mkern-10mu\lambda}_{e}$.

\par In this regard, the second resonant region represented by curve 2 in Fig. 1 are of undoubted interest. It is determined by the second term in square brackets on the right side of Eq. \eqref{9}. For this case, the proton  energy, $\omega_{q}$, is above the lower boundary ($+m_{p}$) of the upper continuum of the Dirac levels. In the same time, the electron energy, $-\varepsilon_{p}$, is negative, and is below the upper boundary ($-m_{e}$) of the lower continuum of the Dirac levels. In this region, the proton momentum $q\geq 2.659\sqrt{m_{e}m_{p}}$. We can estimate the characteristic radius of its motion in the BIC state: $R_{eff} < (2.659\sqrt{m_{e}m_{p}})^{-1}$. As for the $e-$ particle, there are no restrictions on its momentum. Moreover, the electron momentum increases sharply with $q$. It is important that this increase in the momenta $p$ and $q$ does not lead to a change in the energy 
$E=m_{p}+m_{e}+{\mathcal E}$. 

\par Note that for normal bound states, the finding a particle in lower continuum states is not uncommon. For example, in the bound states of the hydrogen, the electron is also characterized by a wave function in the lower continuum. But the probability of being in them is small, of the order of $\simeq \alpha^2$.
This probability increases with the nuclear charge \cite{bib12}.

\par In the second resonant region we have:
\begin{equation}\label{14}
\psi({\bf p},{\bf q};E)=-\frac{\alpha}{8\pi^2\omega_{q}\varepsilon_{p}}\frac{\Lambda_{e}^{-}\Lambda_{p}^{+}}{E+\varepsilon_{p}-\omega_{q}+i\delta}
\int_{D_{k}} \frac{d{\bf k}}{k^2}(1-{\bm \alpha}_{e}{\bm \alpha}_{p})\psi({\bf p}+{\bf k};{\bf q}-{\bf k}).
\end{equation}

In Eq. (14) the integration over $\bf k$ is carried out in the second resonant region $D_{k}$:
\begin{equation}\label{15}
m_{p}+(\gamma+1)m_{e}+\sqrt{m_{e}^2+({\bf p}+{\bf k})^2}=\sqrt{m_{p}^2+({\bf q}-{\bf k})^2}.
\end{equation}

\section{The BIC wave function in the momentum space}
From \eqref{15} we conclude that in the second resonant region the electron is relativistic and its momentum $p>m_{e}$. The proton is non-relativistic. The proton energy can be written 
as $\omega_{q}=m_{p}+\frac{q^2}{2m_{p}}$, and its momentum $q\propto \sqrt{m_{e}m_{p}}$. Then Eq. \eqref{6} is reduced to:   
\begin{equation}\label{16}
\Lambda^{+}_{p} \to 2m_{p}\left(\begin{array}{cc} {1} & 0\\0 & {0} \end{array}\right)_{p}.
\end{equation}
Also, the interaction of the particles through the vector potential can be omitted since  
\begin{equation}\label{17} 
{\bm \alpha}_{e}{\bm \alpha}_{p}\simeq \sqrt{\frac{m_{e}}{m_{p}}}<<1.
\end{equation}

\par Due to the symmetry of the problem, the functions $\psi$ can depend on absolute values of the vectors p and q, and the angle between them $\theta$. So, the function is 
$\psi(p,q;\theta)$.  
In the BIC state, the proton is non-relativistic and, taking into account \eqref{16}, its spin part is the bispinor, in which only the upper spinor is nonzero,
$\left(\begin{array}{cc} {a} \\ {b} \end{array}\right)$ with constant $a$ and $b$. This bispinor is not essential for further consideration. 

\par In Eq. \eqref{14} the operator $\Lambda_{e}^{-}$ given by \eqref{14}, acts on the wave function $\psi({\bf p}+{\bf k};{\bf q}-{\bf k})$. 
The spin operator and the orbital angular momentum operator, each separately do not commute with the operator $\Lambda_{e}^{-}$. For this reason, the wave function cannot have a definite value of the orbital angular momentum and its $z-$projection \cite{bib13}. Therefore, the function can be written as:
\begin{equation}\label{18}
\psi(p,q; \theta)=
\sqrt{\frac{\delta}{\pi}}\frac{1}{E+\varepsilon_{p}-\omega_{q}+i\delta}
\left(\begin{array}{cc} {\frac{1}{\sqrt{3}}v(p,q; \theta)Y_{10}(\frac{{\bf p}}{p})} \\
{\sqrt{\frac{2}{3}}v(p,q; \theta)Y_{11}(\frac{{\bf p}}{p})} \\ 
{u(p,q; \theta)Y_{00}} \\ {0} \end{array}\right)
\end{equation}
The bispinor \eqref{18} contains the spherical harmonics $Y_{00}$, $Y_{10}$ and $Y_{11}$ with the azimuthal quantum number 0 and 1. 

\par From \eqref{18} we obtain  that the two-particle density is determined only in the second resonant region: 
\begin{equation}\label{19}
\vert \psi(p,q; \theta) \vert^2=(\vert v \vert^2+\vert u \vert^2)\delta(E+\varepsilon_{p}-\omega_{q}).
\end{equation}
with the normalization condition
\begin{equation}\label{20}
\int d{\bf p}\int d{\bf q}(\vert v \vert^2+\vert u \vert^2)\delta(E+\varepsilon_{p}-\omega_{q})=1.
\end{equation}

\par Substituting \eqref{18} in \eqref{14} and considering \eqref{16}-\eqref{17}, we obtain:
\begin{equation}\label{21}
\left(\begin{array}{cc} {v({\bf p},{\bf q})} \\ {u({\bf p},{\bf q})} \end{array}\right)= 
\frac{i\alpha}{4\pi \varepsilon_{p}} \int_{D_{k}} \frac{d\bf k}{({\bf k}-{\bf p})^2} 
\delta(E+\varepsilon_{\bf k}-\omega_{{\bf q}+{\bf p}-{\bf k}})
\left(\begin{array}{cc} {(\varepsilon_{p}-m_{e})v({{\bf k};{\bf q}+\bf p}-{\bf k})-pu} \\ 
{(\varepsilon_{p}+m_{e})u({{\bf k};{\bf q}+\bf p}-{\bf k})-pv} \end{array}\right)
\end{equation}
where the second resonant region $D_{k}$ takes now the form:
\begin{equation}\label{22}
m_{p}+(\gamma+1)m_{e}+\sqrt{m_{e}^2+k^2}=\sqrt{m_{p}^2+({\bf q}+\bf p-{\bf k})^2}.
\end{equation}
Note that in this region the principal part 
\begin{equation}\label{23}  
\frac{\cal P}{E+\varepsilon_{\bf k}-\omega_{{\bf q}+{\bf p}-{\bf k}}}=0.  
\end{equation}

\section{Transformation of Eq. \eqref{21}}
Eq. \eqref{21} is the integral equation with the imaginary kernel. So, the functions $v=v_{r}+iv_{i}$ and $u=u_{r}+iu_{i}$ must be complex. Then, the equation \eqref{21} are four coupled integral equations for the four real functions $v_{r,i}(p,q,\theta)$ and $u_{r,i}(p,q,\theta)$. Without loss of generality, we can choose the vector ${\bf p}+{\bf q}$ is directed along the
$z$-axis. Then, Eq. \eqref{21} is rewritten as:
\begin{displaymath}
\left(\begin{array}{cc} {v_{r}(p,q,\theta)} \\ {v_{i}(p,q,\theta)} \\ {u_{r}(p,q,\theta)} \\ {u_{i}(p,q,\theta)} \end{array}\right)= 
\frac{\alpha}{4\pi}\frac{m_{p}}{\varepsilon_{p}\vert {\bf q}+{\bf p} \vert} \int_{0}^{\infty}kdk \int_{0}^{\pi}\sin \theta_{k}
\delta\Bigl(\cos \theta_{k}-\frac{({\bf q}+{\bf p})^2+k^2-q_{*}^2}{2k\vert {\bf q}+{\bf p} \vert}\Bigr)\times
\end{displaymath}
\begin{displaymath}
\int_{0}^{2\pi}\frac{d\phi_{k}}{k^2+p^2-2kp(\cos \theta_{k}\cos \theta_{p}+\sin \theta_{k}\sin \theta_{p}\cos \phi_{k})}\times
\end{displaymath}
\begin{equation}\label{24}
\left(\begin{array}{cc} {-(\varepsilon_{p}-m_{e})v_{i}(k, q_{*}, \theta_{*})+pu_{i}(k, q_{*}, \theta_{*})} \\ {(\varepsilon_{p}-m_{e})v_{r}(k, q_{*}, \theta_{*})-pu_{r}(k, q_{*}, \theta_{*})}
\\ {-(\varepsilon_{p}+m_{e})u_{i}(k, q_{*}, \theta_{*})+pv_{i}(k, q_{*}, \theta_{*})} \\ {(\varepsilon_{p}+m_{e})u_{r}(k, q_{*}, \theta_{*})-pv_{r}(k, q_{*}, \theta_{*})}  \end{array}\right).
\end{equation}
Here $\theta_{*}$ is the angle between the vectors $\bf k$ and ${\bf q}+{\bf p}-{\bf k}$,  
\begin{equation}\label{25}
\cos \theta_{*}=\frac{\vert {\bf q}+{\bf p} \vert \cos \theta_{k}-k}{q_{*}},
\end{equation}
$\phi_{k}$ and $\theta_{k}$ are the azimuthal and the polar angles of the vector $\bf k$, 
\begin{equation}\label{26}
\cos \theta_{k}=\frac{({\bf q}+{\bf p})^2 + k^2 -q_{*}^{2}}{2k\vert {\bf p}+{\bf q} \vert},
\end{equation}
$\theta_{p}$ is the polar angle of the vector $\bf p$; simple geometric considerations give us
\begin{equation}\label{27}
\sin \theta_{p}= \frac{q \sin {\theta}}{\vert {\bf p}+{\bf q} \vert},
\end{equation}
and 
\begin{equation}\label{28}
q_{*}(k)=\vert {\bf p}+{\bf q}-{\bf k} \vert =\sqrt{2m_{p}[(\gamma+1)m_{e}+\varepsilon_{k}]},       
\end{equation}
with $\gamma=1.531$. 

\par After calculations of the integrals over the angles of the vector $\bf k$, Eq. \eqref{24} takes the form:
\begin{displaymath}
\left(\begin{array}{cc} {v_{r}(p,q,\theta)} \\ {v_{i}(p,q,\theta)} \\ {u_{r}(p,q,\theta)} \\ {u_{i}(p,q,\theta)} \end{array}\right)= 
\alpha\frac{m_{p}}{2\sqrt{m_{e}^2+p^2}\vert {\bf q}+{\bf p} \vert} 
\end{displaymath}
\begin{displaymath}
\int_{D_{k}} \frac{kdk}{\sqrt{(k^2+p^2-2kp\cos \theta_{k} \cos \theta_{p})^2-4k^2 p^2 \sin^2 \theta_{k} \sin^2 \theta_{p}}} 
\end{displaymath}
\begin{equation}\label{29}
\left(\begin{array}{cc} {-(\varepsilon_{p}-m_{e})v_{i}(k, q_{*}, \theta_{*})+pu_{i}(k, q_{*}, \theta_{*})} \\ {(\varepsilon_{p}-m_{e})v_{r}(k, q_{*}, \theta_{*})-pu_{r}(k, q_{*}, \theta_{*})}
\\ {-(\varepsilon_{p}+m_{e})u_{i}(k, q_{*}, \theta_{*})+pv_{i}(k, q_{*}, \theta_{*})} \\ {(\varepsilon_{p}+m_{e})u_{r}(k, q_{*}, \theta_{*})-pv_{r}(k, q_{*}, \theta_{*})}  \end{array}\right).
\end{equation}

\section{The effective interaction constant}
\par Given $p\propto m_{e}$ and $k\propto m_{e}$, the integral on the right side \eqref{29} is proportional to $m_{e}$. In the BIC state the proton momentum $q>>p$ and $q\propto \sqrt{m_{p}m_{e}}$.
With account of the latter, the dimensionless factor before the integral in \eqref{29} is proportional to
\begin{equation}\label{30}
\alpha_{eff}=\alpha\sqrt{\frac{m_{p}}{m_{e}}}=0.313.
\end{equation}
This value $\alpha_{eff}$ can be regarded as the effective interaction constant. The fact that $\alpha_{eff}>>\alpha$ is caused by the resonance of the electromagnetic interaction between particles in the BIC state.

\section{The coordinate-space wave function}
\par The momentum-space wave function \eqref{18} satisfies the normalization \eqref{20}. The coordinate-space wave function is defined as:
\begin{equation}\label{31}  
\psi({\bf r}_{e},{\bf r}_{p})=\frac{1}{(2\pi)^3}\int d{\bf q} \int d{\bf p} \psi({\bf p};{\bf q}) 
e^{-i{\bf p}{\bf r}_{e}-i{\bf q}{\bf r}_{p}}.
\end{equation}  
Here ${\bf r}_{e}$ and ${\bf r}_{p}$ are the radius-vectors of the electron and proton, respectively. Taking into account \eqref{31}, the coordinate-space wave function is also normalized:
\begin{equation}\label{32}  
\int d{\bf r}_{e}\int d{\bf r}_{p} \vert \psi({\bf r}_{e},{\bf r}_{p})\vert^2=1.
\end{equation}  

\par The function \eqref{31} can be presented in the form:
\begin{equation}\label{33}
\psi({\bf r}_{e},{\bf r}_{p})=\left(\begin{array}{cc} {f({\bf r}_{e},{\bf r}_{p})} \\ {g({\bf r}_{e},{\bf r}_{p}}
\end{array}\right).
\end{equation}
Here $w$ and $h$ are complex functions which are given by the equation: 
\begin{displaymath}
\left(\begin{array}{cc} {f({\bf r}_{e},{\bf r}_{p})} \\ {g({\bf r}_{e},{\bf r}_{p})}
\end{array}\right)=
N\int d{\bf q} \int d{\bf p} \delta(q^2-2m_{p}((\gamma +1)m_{e}+\sqrt{m_{e}^2+p^2})
\end{displaymath}
\begin{equation}\label{34}
\left(\begin{array}{cc} {v({\bf p},{\bf q})\cos \theta_{p}} \\ {u({\bf p},{\bf q})}
\end{array}\right)
e^{-i{\bf p}{\bf r}_{e}-i{\bf q}{\bf r}_{p}},
\end{equation}  
where $N$ is the normalization factor determined by \eqref{32}.

\section{Details of calculations}
It is convenient to use the dimensionless quantities: $x=p/m_{e}$, $y=q/\sqrt{m_{e}m_{p}}$, $z=k/m_{e}$. Then,  using the notation $\eta=\sqrt{\frac{m_{e}}{m_{p}}}$, 
Eq. \eqref{29} takes the form:
\begin{displaymath}
\left(\begin{array}{cc} {v_{r}(x,y,\theta)} \\ {v_{i}(x,y,\theta)} \\ {u_{r}(x,y,\theta)} \\ {u_{i}(x,y,\theta)} \end{array}\right)= 
\frac{\alpha_{eff}}{2\sqrt{1+x^2}\sqrt{y^2+2\eta xy \cos \theta+\eta^2 x^2}}
\end{displaymath}
\begin{displaymath}
\int_{D_{z}} \frac{zdz}{\sqrt{(z^2+x^2-2xz\cos \theta_{z} \cos \theta_{x})^2-4z^2 x^2 \sin^2 \theta_{z} \sin^2 \theta_{x}}} 
\end{displaymath}
\begin{equation}\label{35}
\left(\begin{array}{cc} {-(\sqrt{1+x^2}-1)v_{i}(z, t_{*}, \theta_{*})+xu_{i}(z, q_{*}, \theta_{*})} \\ {(\sqrt{1+x^2}-1)v_{r}(z, t_{*}, \theta_{*})-xu_{r}(z, t_{*}, \theta_{*})}
\\ {-(\sqrt{1+x^2}+1)u_{i}(z, t_{*}, \theta_{*})+xv_{i}(z, t_{*}, \theta_{*})} \\ {(\sqrt{1+x^2}+1)u_{r}(z, t_{*}, \theta_{*})-xv_{r}(k, t_{*}, \theta_{*})}  \end{array}\right).
\end{equation}
The notations \eqref{25}-\eqref{28} are rewritten as:
\begin{equation}\label{36}
\cos \theta_{*}(x,y,\theta,z)=\frac{1}{t_{*}}\Bigl[\sqrt{y^2+2\eta xy\cos \theta+\eta^2 x^2}\cos \theta_{z}
-\eta z\Bigr],
\end{equation}
\begin{equation}\label{37}
\cos \theta_{z}(x,y,\theta,z)=
\frac{y^2+2\eta xy\cos \theta+\eta (x^2+z^2)-2(\gamma+1+\sqrt{1+z^2})}
{2\eta z\sqrt{y^2+2\eta xy\cos \theta+\eta^2 x^2}},
\end{equation}
\begin{equation}\label{38}
\sin \theta_{x}=\frac{y\sin \theta}
{\sqrt{y^2+2\eta xy\cos \theta+\eta^2 x^2}}.
\end{equation}
and 
\begin{equation}\label{39}
t_{*}=\sqrt{2(\gamma+1+\sqrt{1+z^2})}.
\end{equation}

\par From \eqref{20} we obtain the normalization condition:
\begin{displaymath}
8\pi^2 \int_{0}^{\infty}x^2 (2(1+\gamma+\sqrt{1+x^2}))^{1/2}dx \int_{0}^{\pi}\sin \theta d\theta
\end{displaymath}
\begin{equation}\label{40}
\bigl[\vert v(x,(2(1+\gamma+\sqrt{1+x^2}))^{1/2},\theta) \vert^2+\vert u(x,(2(1+\gamma+\sqrt{1+x^2}))^{1/2},\theta) \vert^2 \Bigr] =1. 
\end{equation}

\par For the coordinate-space wave function we introduce the dimensionless variables,  $s=m_{e}r_{e}$, $t=\sqrt{m_{e}m_{p}}r_{p}$ and 
the angle $\theta_{r}$ between ${\bf r}_{e}$ and ${\bf r}_{p}$. Assuming, without loss of generality, ${\bf r}_{p}$ is directed along the $z-$axis, 
Eq. \eqref{34} is rewritten as:
\begin{displaymath}
\left(\begin{array}{cc} {f(s,t,\theta_{r})} \\ {g(s,t,\theta_{r})} \end{array}\right)=N 
\int_{0}^{\infty}x^2 \sqrt{2(1+\gamma+\sqrt{1+x^2})}dx 
\end{displaymath}
\begin{displaymath}
\int_{0}^{\pi} J_{0}(sx\sin\theta_{r}\sin\theta_{p})e^{-isx\cos \theta_{p}\cos \theta_{r}} \sin \theta_{p}d\theta_{p}
%\end{displaymath}
\int_{0}^{\pi}e^{-it\sqrt{2(1+\gamma+\sqrt{1+x^2})}\cos \theta_{q}}\sin \theta_{q}d\theta_{q}
\end{displaymath}
\begin{equation}\label{41}  
\int_{0}^{2\pi} \left(\begin{array}{cc} {v(x,\sqrt{2(1+\gamma+\sqrt{1+x^2})},\theta)\cos \theta_{p}} \\ {u(x,\sqrt{2(1+\gamma+\sqrt{1+x^2})},\theta)} \end{array}\right)  d\phi_{q}.
\end{equation} 
Here the normalization factor $N$ is determined by the normalization condition, which is easily found from \eqref{33}, $J_{0}$ is the Bessel function of the first kind, and
\begin{equation}\label{42}  
\cos \theta=\cos \theta_{p}\cos \theta_{q}+\sin \theta_{p}\sin \theta_{q} \cos \phi_{q}.
\end{equation} 

\section{Numerical results}
\subsection{Momentum-space BIC wave function}
\par According to \eqref{35}, the momentum-space wave function $\psi(p,q; \theta)$ is determined by the four real functions $v_{r}$, $v_{i}$, $u_{r}$ and $u_{i }$. They depend on three variables $p,q$ and $\theta$, and are interrelated with each other. For the positive binding energy $1.531m_{e}$, the solution of this equation was found numerically using the iteration method. The functions $v_{r,i}(p,q,\theta)$ and $u_{r,i}(p,q,\theta)$ were represented with matrices of the dimension $201\times 201 \times 101$. 

\par It was obtained that for any given angle $\theta$, all these functions represent the single peak located in the same place on the $(p,q)$ plane. The height of the peak changes with the angle. Moreover, it is sign alternating. The heights of peaks are significantly different from zero only for a certain interval of the angles $\theta$.
\begin{figure}[ht]
\centering
\includegraphics[width=16cm]{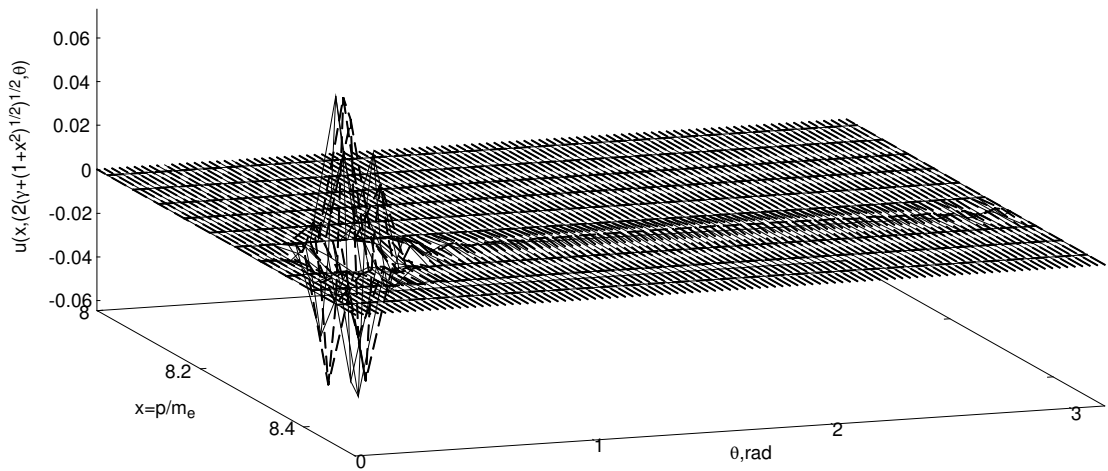}
\caption{The functions $u_{r}(p,q_{*}(p),\theta)$ (solid line) and $u_{i}(p,q_{*}(p),\theta)$ (dashed line). 
\label{f2}}
\end{figure}

\par The value $q_{*}(p)=\sqrt{2m_{p}[(\gamma+1)m_{e}+\varepsilon_{p}]}$ with $\gamma=1.531$ can be considered as the characteristic momentum of the proton in the bound state. 
The point $(p,q_{*})$ does not correspond the maximum of the peaks but falls within the location of the peaks on the $(p,q)$ plane. In Fig. 2 and Fig. 3 we show these four functions depending on the electron momentum $p$ and the angle $\theta$ for the given proton momentum $q_{*}(p)$.
\begin{figure}[ht]
\centering
\includegraphics[width=16cm]{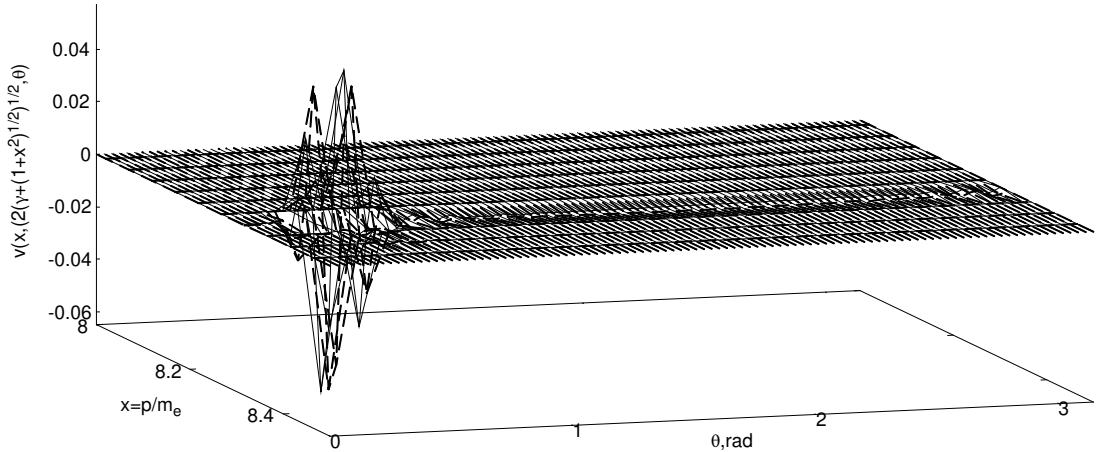}
\caption{The functions $v_{r}(p,q_{*}(p),\theta)$ (solid line) and $v_{i}(p,q_{*}(p),\theta)$ (dashed line). 
\label{f3}}
\end{figure}

\par It can be seen that the main peaks cluster around the angles $\theta \simeq [\pi/18,\pi/4]$. Outside this angular region, these four functions are very small.
Data shown in Fig. 2 and Fig. 3, represent the momentum-space BIC wave function of the electron-proton system.

\par In addition to the state shown in Figs. 2 and 3, one more solution of Eq. \eqref{35} has been obtained. For this second solution, the functions $v_{r}$, $v_{i}$, $u_{r}$ and $u_{i }$  also represent the single peak with variable height depending on the angle $\theta$.
\begin{figure}[ht]
\centering
\includegraphics[width=16cm]{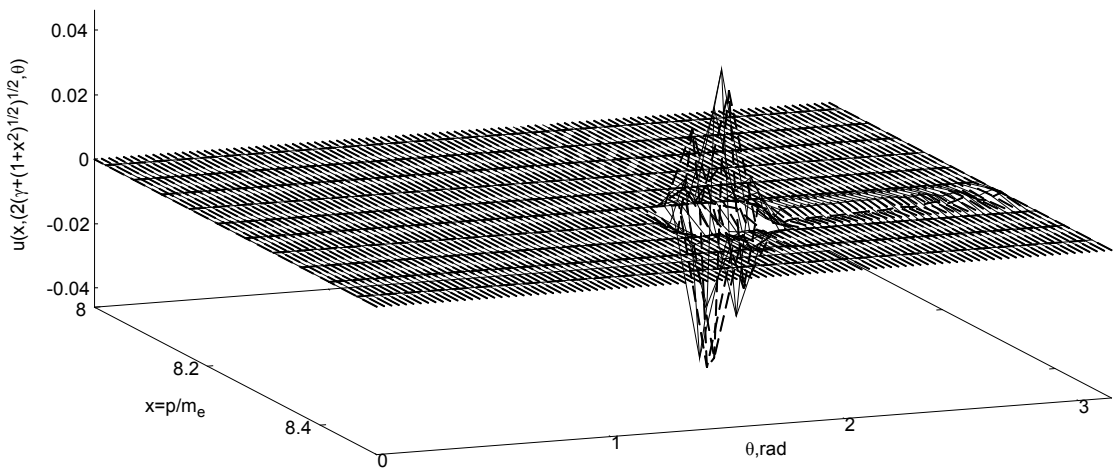}
\caption{The functions $u_{r}(p,q_{*}(p),\theta)$ (solid line) and $u_{i}(p,q_{*}(p),\theta)$ (dashed line) for the second solution of Eq. \eqref{35}. 
\label{f4}}
\end{figure}

\par Results for this second solution are demonstrated in Figs. 4 and 5. They are close to those presented in Fig. 2 and Fig. 3. However, the angular positions of the peaks are shifted
relative to those for the first solution. The shift angle is approximately equal to $\pi/2$.
\begin{figure}[ht]
\centering
\includegraphics[width=16cm]{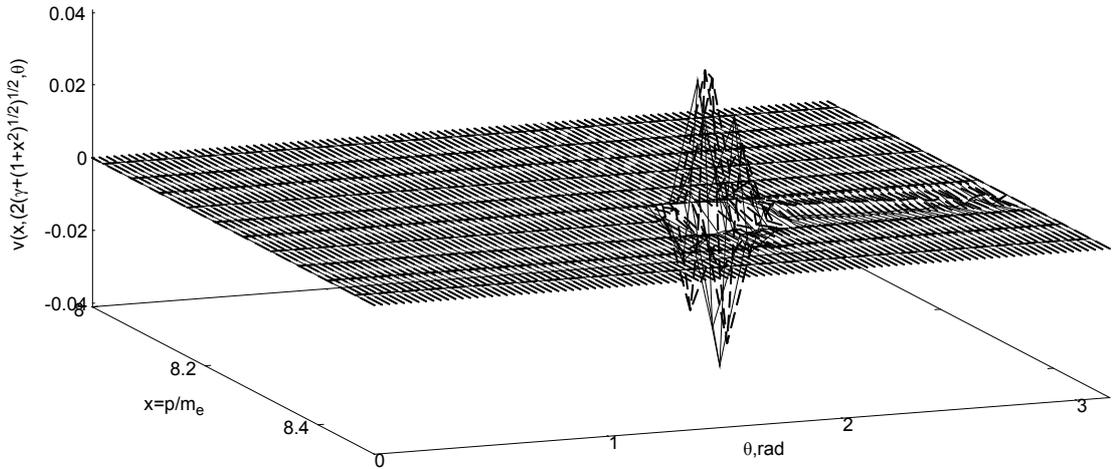}
\caption{The functions $v_{r}(p,q_{*}(p),\theta)$ (solid line) and $v_{i}(p,q_{*}(p),\theta)$ (dashed line) for the second solution of Eq. \eqref{35}. 
\label{f5}}
\end{figure}

\par The two momentum-space BIC wave functions discussed above, are the eigenfunctions of the integral equation \eqref{35} with the eigenvalue $E=m_{p}+2.531m_{e}$. 

\subsection{Coordinate-space BIC wave function}
\par The coordinate-space BIC wave function $\psi(r_{e},r_{p},\theta_{r})$ was calculated from \eqref{41}-\eqref{42}. Since the momentum-space wave functions are normalized, the coordinate-space wave function must also satisfy the normalization condition \eqref{32}. For numerical solution of Eq. \eqref{41} the function $\psi({\bf r}_{e},{\bf r}_{p})$ was replaced by the matrix with dimension $121\times 121\times 17$. Such small dimension is due to the many nested loops involved in numerical procedure.  Of course, this small dimension could affect the accuracy of the calculations. Although the use of matrices with the dimension of $101\times 101\times 17$ does not lead to a significant change in the calculation results.

\par Note that we did not find any difference in the coordinate-space wave functions which have been calculated for these two momentum-space wave functions discussed above.   

\par As it follows from \eqref{16}, in the BIC state the proton is only in the states of the upper continuum. The state of the electron is determined by the bispinor \eqref{33}, which contains two complex functions $f$ and $g$. These functions $f$ and $g$ determine the probabilities of finding the electron in the states of the upper and lower continua, respectively. They are found from Eq. \eqref{41}. It turned out that the function $f$ is very small compared to the function $g$.
\begin{figure}[ht]
\centering
\includegraphics[width=16cm]{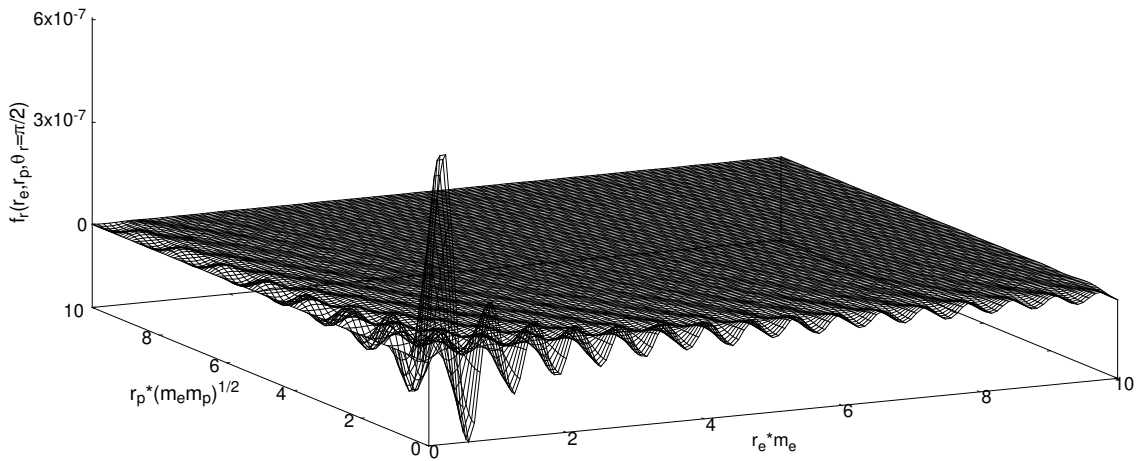}
\caption{The functions $f_{r}(r_{e},r_{p},\theta_{r}=\pi/2)$  
\label{f6}}
\end{figure}
\par For the sake of completeness, the real part of the function $f$ is shown in Fig. 6 for the angle $\theta_{r}=\pi/2$. For this angle, the function $f$ is close to maximal. 
Noteworthy are the very small values of this function. These values are so small that cannot affect the normalization of $\psi$. 

\par Thus, the probability of finding the electron in the states of the upper continuum is negligible. The electron with the probability close to one is in the states of the lower continuum. 
\begin{figure}[ht]
\centering
\includegraphics[width=16cm]{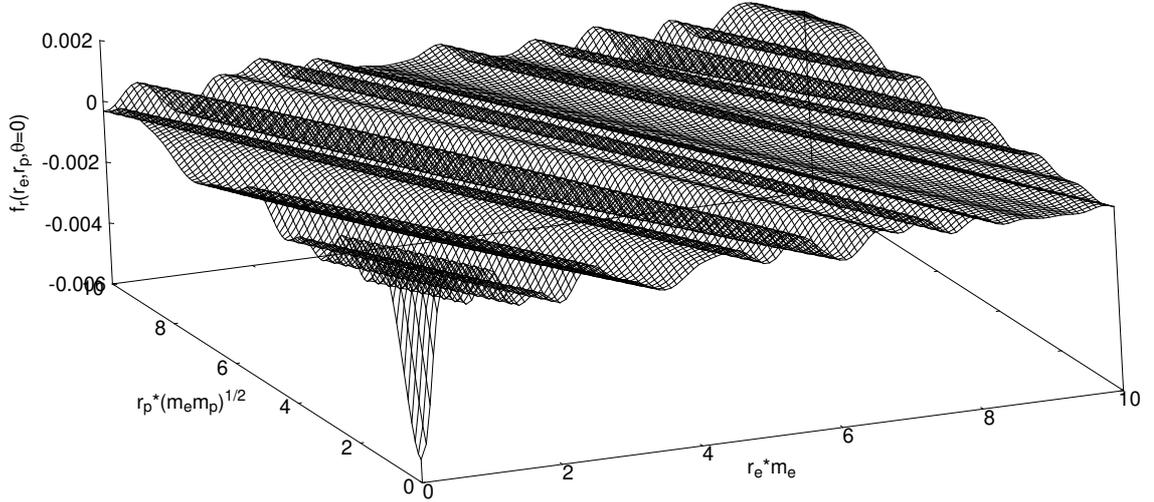}
\caption{The functions $g_{r}(r_{e},r_{p},\theta_{r}=0)$. 
\label{f7}}
\end{figure}

\par  Now the complex function $g=g_{r}+ig_{i}$ is analyzed. It depends on the angle $\theta_{r}$ between the electron (${\bf r}_{e}$) and the proton (${\bf r}_{p}$) radius-vectors. 
According to the calculations, these two functions $g_{r}$ and $g_{i}$ have similar dependencies on $r_{e}$ and $r_{p}$ for a given angle $\theta_{r}$. However, the real part $g_{r}$ has, as a rule, large values compared to $g_{i}$. 

\par The function $g_{r}$ is relatively small at small angles $\theta_{r}\simeq 0$ and for the angles near $\pi$. This function is shown in Fig. 7 for the angle $\theta_{r}=0$.
Attention is drawn to the peak $g_{r}(0,0)$. This peak will be present at all other angles. According to Fig. 7, with the highest probability density, the electron and the proton are near the positions $r_{e}=0$ and $r_{p}=0$.
\begin{figure}[ht]
\centering
\includegraphics[width=16cm]{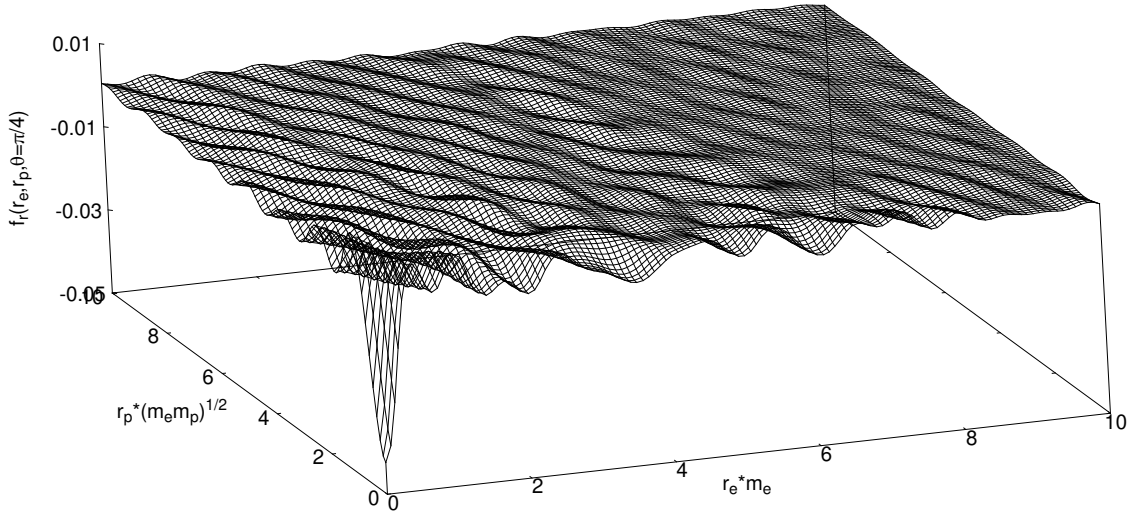}
\caption{The functions $g_{i}(r_{e},r_{p},\theta_{r}=\pi/4)$. 
\label{f8}}
\end{figure}

\par The values of the $g$ function increase with the angle. As shown in Fig. 8, for $\theta_{r}=\pi/4$ the function $g_{r}(r_{e},r_{p},\theta_{r}=\pi/4)$ has also the narrow peak near 
$r_{e}\simeq 0$ and $r_{p}\simeq 0$ . Outside this peak, this function has significantly smaller values. As $r_{e}$ and $r_{p}$ increase, the behavior of $g_{r}$ corresponds to damped oscillations around zero. 

\par The structure of the function $g$ is the same up to the angle $\pi/2$. So, the differences between the functions at $\theta_{r}=\pi/4$ and $\theta_{r}=\pi/2$ are quite insignificant. 
Further, as the angle increases, the situation changes. The central peak is preserved. However, outside the peak, the oscillatory behavior of the $g$ function becomes stronger.
This behavior for the angle $3\pi/4$  is demonstrated in Fig. 9.
\begin{figure}[ht]
\centering
\includegraphics[width=16cm]{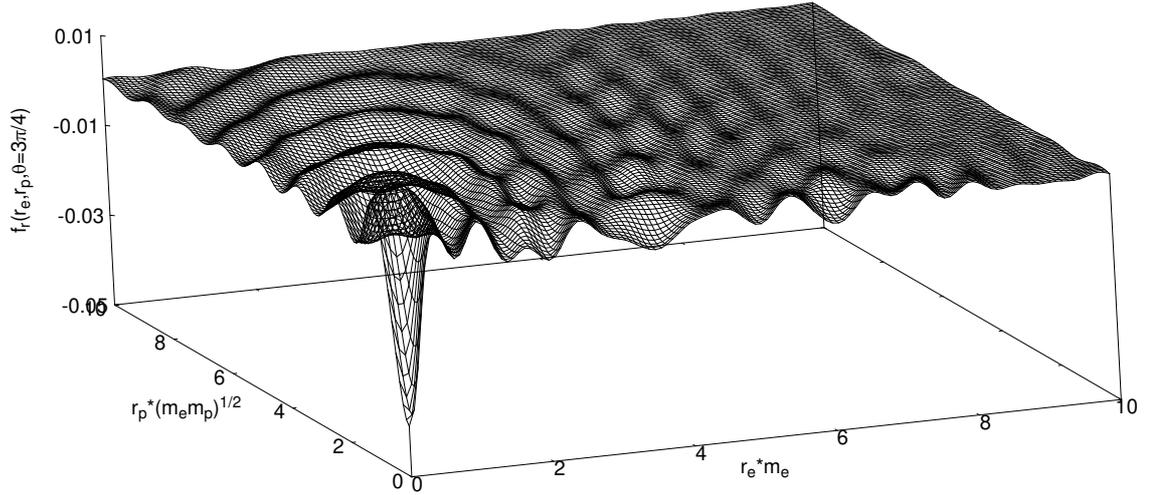}
\caption{The functions $g_{r}(r_{e},r_{p},\theta_{r}=3\pi/4)$. 
\label{f9}}
\end{figure}

The function decrease as the angle $\theta_{r}$ approaches $\pi$. Fig. 10 shows the function $g_{r}$ for $\theta_{r}=\pi$.  
The characteristic values of the function are an order of magnitude smaller than those for the angle  $\theta_{r}=3\pi/4$.
The central peak has been transformed into a deep dip on the undulating surface $g_{r}(r_{e},r_{p}$.
\begin{figure}[ht]
\centering
\includegraphics[width=16cm]{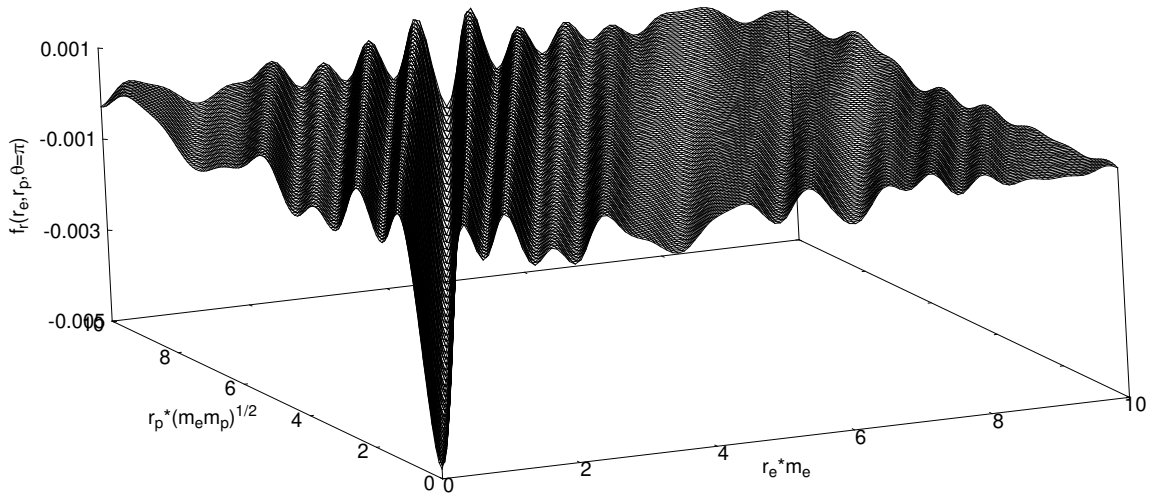}
\caption{The functions $g_{i}(r_{e},r_{p},\theta_{r}=\pi)$. 
\label{f10}}
\end{figure}

The electron-proton BIC state with the positive binding energy can be characterized by the average values of the electron radius $<r_{e}>$ and the proton radius $<r_{p}>$. 
These averages were calculated as follows: 
\begin{equation}\label{43}  
\left(\begin{array}{cc} {<r_{e}>} \\ {<r_{p}>} \end{array}\right)= \int d{\bf r}_{e}\int d{\bf r}_{p} 
\left(\begin{array}{cc} {r_{e}} \\ {r_{p}} \end{array}\right)\vert \psi({\bf r}_{e},{\bf r}_{p})\vert^2.
\end{equation}  
The integration in Eq. \eqref{43} is carried out over the plan $r_{e}\leq 10/m_{e}$ and $r_{p}\leq 10/\sqrt{m_{e}m_{p}}$. According to the obtained data for the wave function, 
which are partially presented in Figs. 6-10, the values of $<r_{e}>m_{e}$ and $<r_{p}>\sqrt{m_{e}m_{p}}$ are very close. We obtained: $<r_{e}>=0.124/m_{e}=48$Fm and 
$<r_{p}>=0.120/\sqrt{m_{e}m_{p}}=1.1$Fm. 

\section{Conclusion}
\par In present work, the theory of BIC states of composite particles was supplemented with the conception of the resonance of interaction between the constituent particles. 
Using the two particle Bethe-Salpeter equation, the resonant regions in momentum space are found with the sharp increase in the electromagnetic interaction between the electron 
and the proton. This increase is so strong that the effective coupling constant is equal to $\alpha \sqrt{m_{p}/m_{e}}=0.313$. Along with correlations in the electron and proton 
motion, this resonance effect determines the confinement mechanism of the composite particle in the BIC state with the positive binding energy of 1.531 of the electron mass.
It was obtained that in the BIC state, the average radius for the electron is equal to 48Fm, and the average radius for the proton is equal to 1.1Fm. 

\par The composite particle from the electron and the proton is a boson with the integer spin, 0 or 1. However, it should be taken into account that $m_{p}>>m_{e}$ and the electron magnetic moment is much greater than the proton magnetic moment. Then, in many experiments, which are considered as direct ones to determine the spin, this composite particle would represent the spin equal to 1/2. For example, experimental setups which are similar to that of Stern and Gerlach, were used in \cite{bib14,bib15}. In these experiments, the particle spin is determined from the splitting of a particle beam when it is passed through a highly inhomogeneous magnetic field. There is no doubt that the splitting of the beam of these composite bosons into two components would certainly be observed. However, the spin of this composite particle is not equal to 1/2. Note that a beam of hydrogen atoms in the ground state would also split into two components, despite the fact that this atom is the boson. As is known, in such experiment, the intrinsic magnetic moment of the electron was established \cite{bib16}. 
Other experiments are also known \cite{bib17,bib18,bib19}, the data of which, for the above reason, do not answer the question what is the spin of the composite particle?

\par Consideration of the neutron as the composite particle also holds for the quark model of baryons. Of course, this model and the approach developed in the present work are completely different. However, as we believe, there is one thing in common. A particle with a certain structure of its constituents exists only in the free state of the composite. It is obvious that in the other composites, for example, in nuclei, the composite particle loses its individuality. In the literature we did not find any experimental works on measuring the form factors of free neutrons. Despite all the complexity, the physical properties of such particles should be studied only when they are free.

\end{document}